\documentstyle[twoside,ijmpa1,psfig]{article}

\def\qed{\hbox{${\vcenter{\vbox{			
   \hrule height 0.4pt\hbox{\vrule width 0.4pt height 6pt
   \kern5pt\vrule width 0.4pt}\hrule height 0.4pt}}}$}}

\newcommand{\kp}{k_{\perp}}

\begin{document}

\runninghead{Note on a Positronium Model from Flow Equations in 
Front-Form Dynamics} 
{Note on a Positronium Model from Flow Equations in 
Front-Form Dynamics} 

\normalsize\textlineskip
\thispagestyle{empty}
\setcounter{page}{1}

\copyrightheading{}			

\vspace*{0.88truein}

\fpage{1}
\centerline{\bf NOTE ON A POSITRONIUM MODEL FROM FLOW EQUATIONS}
\vspace*{0.035truein}
\centerline{\bf IN FRONT-FORM DYNAMICS}
\vspace*{0.3truein}
\centerline{\footnotesize UWE TRITTMANN}
\vspace*{0.015truein}
\centerline{\footnotesize\it Department of Physics, Ohio State University, 
174 W 18th Ave}
\centerline{\footnotesize\it Columbus, OH 43210, USA}
\baselineskip=10pt
\vspace*{-0.0cm}

\vspace*{0.21truein}
\abstracts{
In this note we address the problem of solving for the positronium
mass spectrum. We use front-form dynamics together with the method
of flow equations. For a special choice of the similarity function, the
calculations can be simplified by analytically integrating over the 
azimuthal angle. One obtains an effective Hamiltonian and we solve 
numerically for its spectrum.
Comparing our results with different approaches we find encouraging
properties concerning the cutoff dependence of the results. 
}{}{}


\def\d{\partial}
\def\beq{\begin{equation}}
\def\eeq{\end{equation}}
 
\vspace*{1pt}\textlineskip	
\section{Introduction}
\vspace*{-0.5pt}
\vspace*{0.2cm}

Solving for QCD bound states from first principles remains a yet unsolved
problem. There is some hope of simplifications in the framework of 
front-form dynamics\cite{Review98}, because of a simpler vacuum state. 
Different methods have been developed over the last years
to cope with the stunning problem of constructing an 
effective Hamiltonian out of the 'infinite' canonical QCD Hamiltonian.
The method of flow equations\cite{Wegner94} 
is closely related to the similarity
transformations \cite{WilsonGlazek93}. 
Both methods are based on unitary transformations to block- or 
band-diagonalize the Hamiltonian.
So far no calculations using the fully relativistic and covariant
effective light-cone Hamiltonian were performed in the manner of 
Refs.~\cite{TrittmannPauli97,Krautgaertner92}, where 
first the many body part of the problem is resolved\cite{Pauli98}.
We present numerical results in the similarity flow scheme\cite{Wegner94}
applied to positronium,
with a special choice of the similarity function.
We compare to the results of Ref.~\cite{TrittmannPauli97}, 
and point out the specific virtues of the methods. 


\section{Flow Equations and Similarity Renormalization}

Both the flow equations of Wegner
\cite{Wegner94} and the similarity renormalization of Wilson and G{\l}azek
\cite{WilsonGlazek93} can be retrieved as special cases of the more
general similarity flow framework \cite{Walhout98}. 
The argument goes as follows.
We construct an effective Hamiltonian $H'$ from the canonical 
Hamiltonian $H$ (regularized at some scale $\Lambda$) by a 
similarity transformation,
$H'=UHU^{\dagger}$.
The generator of the transformation is anti-hermitian, $\eta^{\dagger}=-\eta$. 
The Hamiltonian is considered 
a function of the flow parameter $l$, with the bare 
and effective Hamiltonians being $H(l=0)$ and $H(l\rightarrow\infty)$, 
respectively. Its change with respect
to the flow parameter is given by the commutator with the generator
$\eta$ of these unitary transformations
\begin{equation}\label{FlowEqn}
\frac{dH(l)}{dl}=[\eta(l),H(l)].
\end{equation}
\vspace*{-0.2cm}
The goal is to choose an $\eta$ such that the transformed Hamiltonian 
has the form
\begin{equation}\label{form}
H(\lambda)_{ij}=f\left(\frac{E_i-E_j}{\lambda}\right)\hat{H}_{ij}(\lambda).
\end{equation}
The flow parameter is connected to the scale $\lambda$ by 
$l=1/\lambda^2$, and the free particle energies $E_i$ 
are defined by the free Hamiltonian
\begin{equation}\label{base}
H^{(0)}_d|i\rangle=E_i|i\rangle.
\end{equation}
If the so-called similarity function $f(\lambda)$ 
vanishes for vanishing $\lambda$,
then $H(l)_{ij}$ will become a (block) diagonal operator in this limit.
The generator $\eta$ of the unitary transformations is by construction the 
commutator of the diagonal part of the Hamiltonian with its complement,
$\eta=[H_d,V]$.
The two schemes of Wegner and Wilson-G{\l}azek are defined by 
different choices of the generator $\eta$. 
Note that the similarity function $f(\lambda)$ 
is still arbitrary in both schemes.


\section{Positronium Model}\label{SecModel}

A solution of Eq.~(\ref{FlowEqn}) was described in detail in 
\cite{GubankovaWegner98}, so we can be brief here.
One expands the Hamiltonian and the 
generator $\eta$ into a power series in the bare coupling $g$.
With the energies $E_i$ depending on the flow parameter only in second order, 
we can solve the differential equation for the Hamiltonian, Eq.~\ref{FlowEqn} 
in the energy basis defined by Eq.~(\ref{base}) order by order. 
Using the definition
\begin{equation}
V^{(2)}_{ij}(l)=f(l)\hat{V}^{(2)}_{ij}(l), 
\end{equation}
suggested by Eq.(\ref{form}), we get the solutions in the second order
\begin{eqnarray}\label{Hamilton}
H^{(2)}_{d,ij}(l)&=&H^{(2)}_{ij}(l=0)+
\int_{0}^{l}
[\eta^{(1)},H^{(1)}]^{(d)}_{ij}(l')d l',\\
\hat{V}^{(2)}_{ij}(l)&=&\hat{V}^{(2)}_{ij}(l=0)+
\int_{0}^{l}f(l')
[\eta^{(1)},H^{(1)}]^{(V)}_{ij}(l')dl',\label{Hamilton2}
\end{eqnarray}
where the superscripts $(d)$ and $(V)$ denote the diagonal and the particle
number changing part, respectively.
We are supposed to take the bare cutoff $\Lambda$, which defines the terms at 
$l=0$, to infinity.

To evaluate Eqs.~(\ref{Hamilton}) 
and (\ref{Hamilton2}) and to obtain the matrix elements of the Hamiltonian, 
we choose the particle number 
conserving part of the Hamiltonian as diagonal, and 
we reduce the particle number 
violating blocks of the Hamiltonian to zero with the flow equations.
This solves {\em en passant} also the multi-particle problem. Instead of having
to truncate the Fock space {\em a l\`a} Tamm-Dancoff, 
we are now dealing with isolated blocks of definite particle number. 
The effective matrix elements are thus obtained by the process
\begin{equation}
V_{\rm eff}=\lim_{\lambda\rightarrow 0} \left(V^{gen}+V^{PT}\right).
\end{equation}
The interaction can be read off the structure of Eqs.~(\ref{Hamilton}) and
(\ref{Hamilton2}):
the part without the integral ($V^{PT}$) is the one obtain by usual 
perturbation theory. The
second one ($V^{gen}$) is generated by the flow of the Hamiltonian. 
In other words, by reducing the off-diagonal matrix elements, we are inducing
changes on the diagonal.

To calculate the matrix elements,
one evaluates the associated diagrams, applying light-cone
perturbation theory \cite{LepageBrodsky80}. The electron [positron] 
momenta and currents are 
$l_e^{\mu}=(k'_e-k_e)^{\mu}  
[l_{\bar{e}}^{\mu}=(k_{\bar{e}}-k'_{\bar{e}})^{\mu}]$, and
$j(l_e)^{\mu}=\bar{u}(k'_e)\gamma^{\mu}u(k_e)
[j(l_{\bar{e}})^{\mu}=\bar{u}(k'_{\bar{e}})\gamma^{\mu}u(k_{\bar{e}})]$, 
respectively.
Here, $k_{\bar{e}}^{\mu}$ and $k_{\bar{e}}^{'\mu}$ are the 
electron momenta before and after the interaction.
For the numerical calculations we use relative coordinates: 
$x=\frac{p^+}{P^+}$ is the longitudinal momentum fraction, and
$\vec{k}_{\perp}$ is the transverse momentum.
We obtain 
\begin{eqnarray}\label{Veff}
V^{gen}_{\lambda}&=&\frac{j(l_e)^{\mu}j_{\mu}(l_{\bar{e}})}
{\tilde{\Delta}_1}
\int_{\lambda}^{\Lambda}\frac{df_{\lambda'}(\Delta_1)}{d\lambda'}
f_{\lambda'}(\Delta_2)d\lambda' + (1\leftrightarrow 2)\nonumber \\
V^{PT}_{\lambda}&=&
\frac{j(l_e)^{\mu}j_{\mu}(l_{\bar{e}})}{\cal D}f_{\lambda'}(\Delta_1)
f_{\lambda'}(\Delta_2)+\frac{j^+(l_e)j^+(l_{\bar{e}})}{|x-x'|}(T^*-\omega).
\end{eqnarray}
Here, 
$T^*=\frac{1}{2}\left(k_e+k_{\bar{e}}\right)^2+
     \frac{1}{2}\left(k'_e+k'_{\bar{e}}\right)^2$
is the average kinetic energy before and after the interaction and $\omega$
is one of the (unknown) eigenvalues of the full Hamiltonian. The latter
ambiguity is no problem in the formalism considered here, because 
the perturbative term vanishes when the scale $\lambda$ goes to zero. 
The energy denominator is given by
\begin{equation}
{\cal D}=|x-x'|(T^*-\omega)-\frac{1}{2}(l_e^2+l_{\bar{e}}^2),
\end{equation}
and we used the definition $\Delta_i={\tilde{\Delta}_i}/({x-x'})$, where
\begin{eqnarray}
\tilde{\Delta}_1={m^2_f}\frac{(x-x')^2}{xx'}
     +\frac{x'}{x}k_{\perp}^2+\frac{x}{x'}k_{\perp}^{'2}
     -2k_{\perp}k'_{\perp}\cos(\varphi-\varphi'),
\end{eqnarray}
and $\tilde{\Delta}_2=\tilde{\Delta}_1(x\rightarrow 1{-}x, 
x'\rightarrow 1{-}x')$.
The similarity function $f_{\lambda}(\Delta)$ 
in the interaction is still at our disposal.
In the electron-positron sector we have the integral equation
\begin{eqnarray}\label{TDEquation}
0&=&\left(\frac{m^2 + \vec{k}^2_{\perp}}{x(1-x)}- M_n^2\right)
\psi_n(x,\vec{k}_{\perp};\lambda_1,\lambda_2)\\
&+&\frac{g^2}{16\pi^3}
\sum_{\lambda'_1,\lambda'_2}\int_D dx'd^2\vec{k}'_{\perp}
\langle x,\vec{k}_{\perp};
\lambda_1,\lambda_2|V_{\rm eff}|x',\vec{k}'_{\perp};
\lambda'_1,\lambda'_2\rangle
\psi_n(x',\vec{k}'_{\perp};\lambda'_1,\lambda'_2).\nonumber
\end{eqnarray}
The cutoff enters into the problem via the definition of the integration
domain. It is restricted by the Lepage-Brodsky cutoff on the kinetic energies
$(m^2+\vec{k}^2_{\perp})/{x(1-x)}\le$ $\Lambda^2+4m^2$.
To simplify the numerical calculations we integrate out the 
azimuthal angle in the problem 
($\vec{k}_{\perp}=\kp e^{i\varphi}$) by substituting it by the 
discrete quantum number $J_z$, {\em cf.~} 
Refs.~\cite{Krautgaertner92,TrittmannPauli97}. 
To be able to perform the corresponding 
integral analytically, we have to choose a special similarity function
\begin{equation}\label{special}
f_{\lambda}(\Delta)=\theta(\lambda^2-|\Delta|).
\end{equation}
This sharp cutoff leads to difficulties in the calculations, namely 
the collinear singularity is not canceled exactly
anymore. 
The scale integrals, however, become very simple 
in the limit $\lambda\rightarrow 0$, yielding a theta function,
$\theta(\Delta_i-\Delta_j)$.

\begin{figure}
\centerline{
\psfig{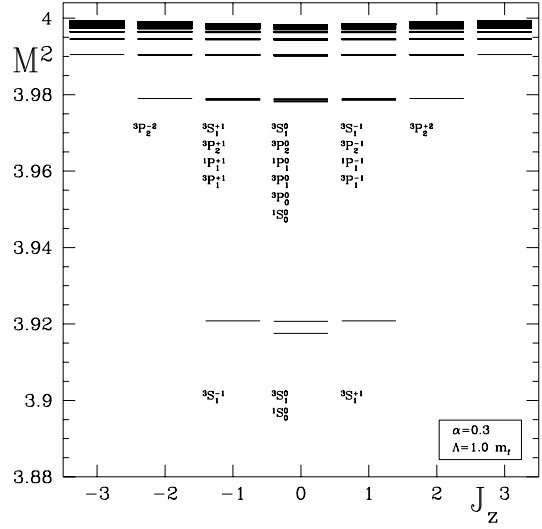}}
\vspace*{0.3cm}
\caption{Positronium spectrum($\alpha=0.3, 
\Lambda=1.0\, m, N_1=N_2=21$):
bound state masses $M^2_n$ in units of the electron mass $m^2_f$
in different $J_z$ sectors.  
} 
\label{yrast}
\end{figure}

The actual matrix elements, integrated analytically over the azimuthal angle
differ only by a re-definition
of the spin-dependent function $Int(n)$ from the matrix elements of 
the Pauli ansatz, {\em cf.~e.g.}~Ref.~\cite{TrittmannPauli97b}. 
This makes it possible to use the numerical techniques and the computer code 
of Ref.~\cite{TrittmannPauli97b} to calculate the positronium spectrum. 
The matrix elements used in the present note
have an additional singular spin-independent part. We argue that this is an
artifact of the choice of the similarity function and will omit it in the
numerical calculations. We comment on the justification of this step
below. 
The interaction used here is the same as listed in Appx. F of 
Ref.~\cite{TrittmannPauli97b}, with the following changes. 
\begin{eqnarray*}
{G}_1(x,{k}_{\perp};x',{k}_{\perp}')&\rightarrow&G_1 
        +\frac{|a_1-a_2|}{(x-x')^2}\mbox{\bf Int}(|1-n|)\delta_{n,1}\\
{G}_2(x,\kp;x',\kp')&\rightarrow&G_2
      +\frac{|a_1-a_2|}{(x-x')^2}\mbox{\bf Int}(|n|)\delta_{n,0},
\end{eqnarray*}
and in all $G_i$ the function $Int(n)$ is replaced by
\begin{equation}
\mbox{\bf Int}(n)=\theta(a_1-a_2)Int_1(n)+\theta(a_2-a_1)Int_2(n),
\end{equation}
where
\begin{equation}
Int_i(n) := -\frac{\alpha}{\pi}(a_i^2-4k_{\perp}^2 k_{\perp}^{'2})^{(n-1)/2}
          \left(\frac{a_i({a_i^2-4k_{\perp}^2 k_{\perp}^{'2}})^{-1/2}-1}
	{2\kp\kp'}\right)^n.
\end{equation}
Here
\begin{eqnarray}
a_1&=&{m^2_f}\frac{(x-x')^2}{xx'}
     +k_{\perp}^2+k_{\perp}^{'2}
   -(x-x')\left[\frac{k_{\perp}^{2}}{x}-\frac{k_{\perp}^{'2}}{x'}\right],
\end{eqnarray}
and $a_2=a_1(x\rightarrow 1-x, x'\rightarrow 1-x')$.
Note that the analogue of the latter expressions in the calculations with
the Pauli ansatz \cite{TrittmannPauli97}, is their average $a=(a_1+a_2)/2$.


\section{Numerical results}\label{SecNum}

The Hamiltonian matrix elements were derived by applying the flow equation
scheme to the positronium problem in front-form dynamics.
Contrary to preceding work, at this point of the calculations we use 
a non-perturbative method to extract the spectrum of this Hamiltonian rather
than light-cone bound state perturbation theory.
We have to solve the eigenvalue problem
$H_{LC}|n\rangle=M_n^2|n\rangle$
or, equivalently, the integral equation, Eq.~(\ref{TDEquation}).
We use the algorithm set up in Ref.~\cite{TrittmannPauli97b}. 
For details of the
calculations and numerical methods applied, see there. To be able to trace
possible violations of rotational invariance (a non-trivial issue in 
front-form dynamics), we chose to work with an unphysically large 
coupling constant
$\alpha=0.3$. This is possible, because the integral equation is an algebraic
function of the coupling.


\begin{figure}[t]
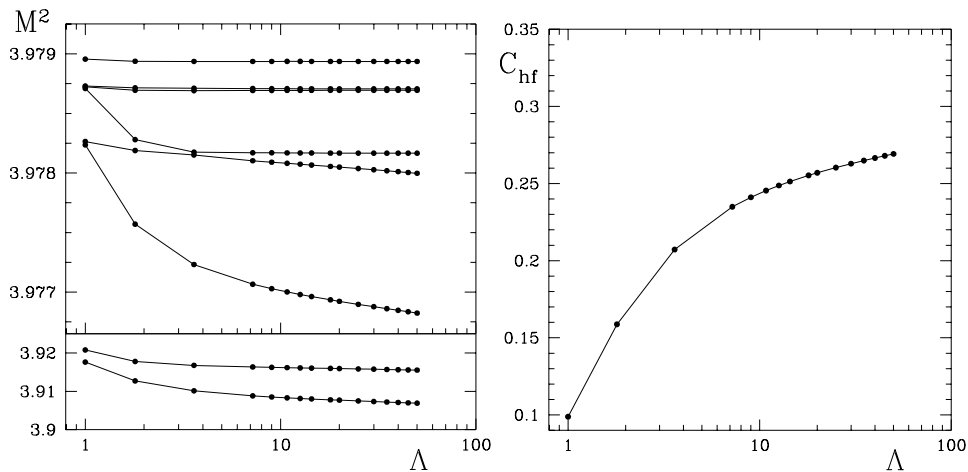

\centerline{
\psfig{figure=lambda_J0_all.epsi,width=6.4cm,angle=0}
\psfig{figure=HFS.epsi,width=6.2cm,angle=0}}
\protect\caption{Cutoff 
dependence: (a) Eigenvalues. 
Below: triplet (upper curve) and singlet (lower curve) ground state. 
Above: first excited states $(n{=}2).$
(b) Hyperfine structure coefficient 
$C_{hf}$. The cutoff is given in units of the electron mass
($\alpha{=}0.3, N_1{=}25, N_2{=}21$).}
\label{lambdaEV}
\end{figure}

The results of the computations are compiled in Fig.~\ref{yrast}.
We get the expected Bohr spectrum, the hyperfine splitting
and even the correct multiplet structure. 
The multiplets are (almost) degenerate. We find exponential 
convergence of the eigenvalues when approaching the continuum limit 
($N\rightarrow\infty$), as in Ref.~\cite{TrittmannPauli97}. 

The eigenvalues depend weakly on the cutoff $\Lambda$, 
{\em cf.}~Fig.~\ref{lambdaEV}(a), except for the singlets. 
From analytic arguments, 
we expect the eigenvalues to diverge logarithmically because
we used flow equations derived up second order only. However, as we shall see,
the coefficient of the hyperfine splitting is very well described at moderate
values of the cutoff. We therefore compare our eigenvalues to results of 
equal time perturbation theory at $\Lambda=50m_f$.
The eigenvalues are listed in Table \ref{TableEVs} and agree 
{\em cum grano salis} with the 
known results. Note, however, that perturbation theory might be not very
reliable at this large coupling. 
\begin{table}[t]
\centerline{
\begin{tabular}{r|ccc||r|ccc}\hline\hline
\rule[-3mm]{0mm}{8mm}$n$ &$M^2_{n}(J_z{=}0)$ & $M^2_{PT}$ &
 $\Delta M^2$ & $n$ &$M^2_{n}(J_z{=}0)$ & $M^2_{PT}$ &
 $\Delta M^2$\\ \hline
 1 & $        3.906908$ & $        3.900002$ & $        0.6906$ & 
10 & $        3.990189$ & $        3.989675$ & $        0.0513$ \\ 
 2 & $        3.915533$ & $        3.910673$ & $        0.4860$ & 
11 & $        3.990239$ & $        3.989825$ & $        0.0413$ \\ 
 3 & $        3.976824$ & $        3.975860$ & $        0.0964$ &
12 & $        3.990345$ & $        3.989875$ & $        0.0470$ \\ 
 4 & $        3.977998$ & $        3.976533$ & $        0.1465$ & 
13 & $        3.990368$ & $        3.989875$ & $        0.0492$ \\ 
 5 & $        3.978166$ & $        3.977037$ & $        0.1129$ & 
14 & $        3.990423$ & $        3.989945$ & $        0.0478$ \\ 
 6 & $        3.978694$ & $        3.977206$ & $        0.1488$ & 
15 & $        3.990432$ & $        3.989945$ & $        0.0487$ \\ 
 7 & $        3.978705$ & $        3.977206$ & $        0.1500$ & 
16 & $        3.990441$ & $        3.989955$ & $        0.0486$ \\ 
 8 & $        3.978936$ & $        3.977441$ & $        0.1495$ & 
17 & $        3.990506$ & $        3.989955$ & $        0.0551$ \\ 
 9 & $        3.989893$ & $        3.989476$ & $        0.0417$ & 
18 & $        3.990519$ & $        3.989984$ & $        0.0536$ \\ 
\hline\hline
\end{tabular}}
\caption{Positronium spectrum ($\alpha{=}0.3,\Lambda{=}50\, m,N_1{=}N_2{=}21$):
present results ($M^2_{n}$), perturbation theory ($M^2_{PT}$), and  
difference ($\Delta M^2{=}M^2_{n}{-}M^2_{PT}$) in percent.}
\label{TableEVs}
\end{table}

A test for the theory is the value of the coefficient of the 
hyperfine splitting,  
$C_{hf}={(M_{triplet}-M_{singlet})}/{m\alpha^4}$.
The plot of $C_{hf}$ versus the cutoff, Fig.~\ref{lambdaEV}(b), is 
encouraging: we obtain a smooth curve, converging for large cutoffs
to a value
$C_{hf} 
\stackrel{\Lambda\rightarrow\infty}{\longrightarrow} 0.2825.$
From equal time perturbation theory we would expect a value between 
$1/3$ (order $\alpha^4$) and $0.2379$ 
(order $\alpha^6 \log \alpha$), which is exactly what we obtain.

Let us focus on violations of rotational invariance, 
{\em cf.}~Ref.~\cite{TrittmannPauli97}. 
We fitted the cutoff dependence of the discrepancy of corresponding 
eigenvalues of different $J_z$ to a polynomial in $\log \Lambda/m_f$.
The (negative) linear coefficient is smaller than $10^{-3}$,
and the constant term is ca.~$10^{-4}$. This means that
the discrepancy between the triplet levels is 1\%
of the relevant (hyperfine splitting) scale at a cutoff of one fermion mass,
and rises to roughly 10\% when $\Lambda=18m_f$.
This suggests that to obtain full rotational invariance, one has to go
to higher orders in the derivation of the flow equations.

The original idea of the flow-equation approach 
implies the use of a smooth similarity function to
cancel the collinear singularity completely. We had to omit 
the spin independent singular part by hand to be able to perform the 
integration of the numerical counterterms. We argue that the singular part
is an artifact of the choice of the similarity function. Indeed, it can be 
shown that the singular part of the matrix elements vanishes for the 
standard flow equation similarity function.
There is a danger that this choice 
might change the eigenfunctions significantly. 
We found, however, that the wavefunctions
are almost identical with those of Ref.~\cite{TrittmannPauli97}.


\section{Discussion}

We presented the full positronium spectrum and 
wavefunctions using flow equation techniques
in all sectors of the angular momentum $J_z$. The results
are encouraging especially when looking at their cutoff dependence. 
Quite in general we find a weak, logarithmic, smooth dependence on the cutoff. 
The comparison to results 
of equal time perturbation theory gives good agreement. 
With respect to its cutoff related properties, the flow equation scheme
seems to work
better than the model of Ref.~\cite{TrittmannPauli97}, inspired by the 
method of iterated resolvents. This is expected by construction of the
methods and new ideas have been proposed to deal with renormalization issues
in the latter method \cite{Pauli98}. 
The inclusion of the annihilation channel would be a straightforward 
calculation, cf.~\cite{Trittmann97c}. 
We found that rotational invariance is obeyed to a large extent 
on the numerical level 
and withstood therefore from implementing this channel here. 
It remains to be investigated how the singular terms influence the spectrum.
We argued that they are relics of the chosen similarity function and omitted 
them. The numerical results seem to support this claim.
In general, it would be interesting to calculate the effective 
Hamiltonian to a higher order in the bare coupling constant.
This would be useful to find out about the structure of the generated 
(irrelevant) operators and also to test
the improved cutoff dependence of the results. 

\vspace{0.2cm}

{\bf Acknowledgments:}
This work was done while the author was a Minerva fellow at the Weizmann 
Institute, Israel. 
The author is indebted to E.~Guban\-kova for enlightening 
discussions about her analytic calculations.


\end{document}